\documentstyle[aps,prc]{revtex}

\begin{document}
\author{P. Balenzuela and C.O.Dorso}
\title{Information entropy in fragmenting systems}
\date{\today}
\maketitle

\begin{abstract}
The possibility of facing critical phenomena in nuclear fragmentation is a
topic of great interest. Different observables have been proposed to
identify such a behavior \cite{stauffer,dorbona,bauer,campi,elliot}, in
particular, some related to the use of information entropy as a possible
signal of critical behavior \cite{Ma}. In this work we critically examine
some of the most widespread used ones comparing its performance in bond
percolation and in the analysis of fragmenting Lennard Jones Drops.
\end{abstract}

\vspace{2cm}

When nuclei collide at energies of the order of hundreds of MeVs, the
resulting highly excited system breaks up into several medium-size
fragments. This phenomenon is usually referred to as multifragmentation and
attracts a lot of attention in the nuclear community because it is
conjectured that the fragmenting system might undergo a phase transition.
In particular the possibility of facing critical
behavior (originally triggered by the seminal work of the Purdue Group \cite
{purdue}) has induced the development of different signatures to
characterize such a behavior from the analysis of the observed mass spectra.
Starting with the work of Bauer \cite{bauer} and Campi \cite{campi} the
percolation model has emerged as a quite useful test-bed for the analysis of
the tools developed to study nuclear multifragmentation.

In \cite{Ma} the information entropy $(S_{1})$ has been proposed as a
signature of critical behavior and two systems has been analyzed: a lattice
gas model and a molecular dynamics model.

In this work we test the information entropy together with other proposed
signatures of critical behavior in a system (percolation on a simple cubic lattice) 
that it is known to undergo a second order phase transition. For this model, the
behavior near the critical point (including the critical exponents) is well
known. Besides the information entropy, we also analyze the normalized variance 
of the size of the maximum fragment ($NVM$) \cite{dorbona}, the second moment of the 
distribution ($M_{2}$) \cite{stauffer}, the best fitting power law \cite{elliot} and
minimum $\tau $ \cite{Ma,bruno}.

We have calculated the above mentioned signatures from the mass
distributions resulting from a bond percolation process on two simple cubic
lattices  of sizes $30\times30\times30$ and $6\times6\times6$. It is known that for 
the larger
one, the finite size effects are negligible, meanwhile for the smaller one
(whose size is of the order of the one corresponding to nuclear system's size) the 
finite size effects are important \cite{elliot_perco}.

Once the various criticality signatures have been tested we use them to
analyze excited Lennard-Jones drops undergoing fragmentation. We focus on
whether the non-equilibrium process of fragmentation can be characterized as
a critical phenomenon.

\smallskip In percolation, the number of clusters of size $A$ per lattice
site, $n_{A}$, at a given bond activation probability, $p$, was proposed in 
\cite{stauffer} to be given by :

\begin{equation}
n_{A}(p)=q_{0}A^{-\tau }f(z) ,   \label{eq:fdm}
\end{equation}

where $q_{0}$ is a normalization factor, $\tau $ a critical exponent and $%
f(z)$ a scaling function. This equation shows that $n_{A}(p)/q_{0}A^{-\tau }$
depends on $p$ and $A$ via the combination 
$z=((p-p_{c})/p_{c})A^{\sigma}=\epsilon A^{\sigma}$.
Thus it is clear that at the
critical point $n_{A}(p_{c})=q_{0}A^{-\tau }f(0)$ with $f(0)=1$. All this is
valid for not too small fragments and infinite systems for which the
percolating cluster has been removed. In finite systems we won't get a pure
power law because finite size effects introduce distortions. Nevertheless,
there exists a range of masses (depending of the size of the system) for
which the mass spectra presents a power law like behavior, as can be seen in
figure (1f).

There are two main methods to find the critical point from the mass spectra.
The first one consists in fitting the mass spectra with a simple power law,
with two independent parameters, $q_{0}$ and $\tau $, and then look for the
value of $p$ (or the quantity taken as control parameter) for which the
exponent $\tau $ is minimum \cite{Ma,bruno}.

The second one was developed in \cite{elliot} and takes care of the
constraint imposed by the normalization, i.e. $q_{0}=q_{0}(\tau )$. This
dependence can be found by analyzing the first moment of the cluster
distribution at the critical point \cite{stanley}:

\begin{equation}
M_1(\epsilon=0)=\sum_A n_A(\epsilon=0)A=q_0\sum_A A^{1-\tau}=1 ,
\label{eq:norm}
\end{equation}

with the sum running over all clusters. Equation (\ref{eq:norm}) shows
that the normalization constant, $q_{0}$, depends on $\tau $ via a Riemann $%
\zeta $ function:

\begin{equation}
q_0=\frac{1}{\sum_A A^{1-\tau}}  .  \label{eq:q0}
\end{equation}

Therefore, if we take into account the dependence of $q_{0}$ with $\tau $
via equation (\ref{eq:q0}), we have to fit the cluster distribution  
with a single parameter power law. 
In this way, the critical point corresponds to the best fitted spectra, i.e.
minimum $\chi ^{2}$.

The critical point can also be found by looking for the maximum
(divergence in an infinite system) of the moments of the distribution \cite
{stauffer} or of the normalized variance of the size of the maximum fragment
(NVM).

The moments of the cluster distribution, $M_k$, are defined as:

\begin{equation}
M_k=\sum_A A^k n_A  ,
\end{equation}

where $n_{A}$ is the cluster distribution.

The normalized variance of the size of the maximum fragment (NVM) \cite
{dorbona} is defined as:

\begin{equation}
NVM=<\frac{<A_{max} - <A_{max}>>^2}{<A_{max}>}> ,
\end{equation}

where $A_{max}$ is the largest fragment of a given event and $<...>$ is an
average over an ensemble of events at a given probability $p$.

Finally, the information entropy, as proposed in \cite{Ma}, is defined as:

\begin{equation}
S_1=-\sum_A p_{A}\ln (p_{A}) ,  \label{eq:S1}
\end{equation}

where $p_{A}$ is the probability of detecting a fragment of mass $A$ defined
as $p_{A}=N_{A}/N_{t},$ with $N_{A}$ the number of fragments of size $A$
detected in a set of experiments and $N_{t}=\sum_{A}N_{A}$ the total number
of fragments detected in the same set of experiments. This magnitude is
referred to as $S_{1}$ because it belongs to the family of generalized Renyi
entropies defined by \cite{dorbo}:

\begin{equation}
S_q=\left\{ 
\begin{array}{ll}
-\sum_p p ln(p) & q=1 \\ 
\frac{-1}{q-1}ln(\sum_p p^q) & q \ne 1 .
\end{array}
\right.
\end{equation}

In figure (1) we show the results of our calculations for percolation three 
dimensional lattices of two different sizes specified above.
The $NVM$ in figure (1a) , $M_{2}$ in figure (1b), $\chi^{2}$ in figure (1c) (the
minimum of $\chi ^{2}$ denotes the best fitting power law with one-parameter
fit), the $\tau _{eff}$ in figure (1d) (corresponding to a two-parameter fit of
the cluster distribution. The minimum value of $\tau $ would correspond to
the critical point) and $S_{1}$ in figure (1e).

We can see that for the mass spectra corresponding to a lattice of size 
$30\times30\times30$ (filled symbols of figure (1)) the signals $NVM$, $M_{2}$ 
and $\chi ^{2}$ give the same
critical probability ($p_{c}=0.25\pm 0.01$). The extremes are very sharp
and the critical probability is the one corresponding to the infinite system limit,
indicating that the finite size effects are negligible. On the other hand,
the maximum of the information entropy is not as sharp and it is also
shifted (at $p=0.225\pm 0.015$) with respect to the extreme of the other
signatures, showing that it is not a good signature of the critical point.
Finally, we can see that when we fit the cluster distribution with a power
law with two independent parameters the exponent $\tau $ is minimum at
$p=0.24\pm 0.01$, slightly shifted from the other signals (but with
overlapping error bars) and with a minimum less pronounced than when we fit
the mass spectra with $\tau $ and $q_{0}(\tau )$ (minimum $\chi ^{2}$).

For the smaller ($6\times6\times6$) lattices (open symbols in figure (1)),
we see that the peaks of the various signatures analyzed are slightly
shifted from one another but with overlapping error bars, even for the case
of the information entropy. The best fitted mass spectra according to the
condition given by equation (\ref{eq:q0}) is at $p=0.31\pm 0.01$ (figure (1c),
the $NVM$ is maximum at $p=0.30\pm 0.01$ (figure (1a)), the $M_{2}$ at $p=0.29\pm
0.01$ (figure (1b)) and $S_{1}(p_{i})$ at $p_{i}=0.27\pm 0.05$ (figure(1e)).

As we can see, when the system is small, finite size effects can introduce
distortion and smear out the signature of the critical point. On the other
hand, $S_{1}$ is shifted with respect the critical point when finite size
effects are negligible.

In order to understand the behavior of the information entropy in this kind
of systems, we have chosen to generate a family of clusters distributions
according to equation (\ref{eq:fdm}) with the critical exponents corresponding to
percolation in an infinite lattice in three dimensions. In this case, the
normalization constant $q_{0}$ is taken as the inverse of the zeta Riemann
function evaluated at $\tau =2.18$ and the scaling function $f(z)$ is the
one fitted in \cite{elliot_f}. We have also chosen $p_{c}=0.25$ i.e. the
infinite size value. These cluster distributions have the same functional
form as the ones corresponding to a percolation problem, but without the
percolating cluster and without finite size effects.

The information entropy can then be calculated by replacing $n_{A}(p)$ in
the definition of $S_{1}$ (equation (\ref{eq:S1})) by its expression according to 
equation (\ref{eq:fdm}) .
The results of this calculation are plotted in figure (2c) and we can see
that the maximum in the entropy does not correspond to the pure power law
distribution ($p_{c}=0.25$ in this case) as was claimed in \cite{Ma} but to
the distribution corresponding to $p=0.22$ showed in figure (2d). We can
check this result by calculating the derivative of $S_{1}$ against $\epsilon
=(p-p_{c})/p_{c}$:

\begin{equation}
\frac{d S_1}{d \epsilon} = q_0^2 \sum_{s_1 s_2} s_1^{(-\tau)} s_2^{(-\tau)}
s_1^{\sigma} f^{\prime}(s_1,\epsilon) f(s_2,\epsilon)
log[(s_1^{-\tau}f(s_1,\tau))/(s_2^{-\tau}f(s_2,\tau))] ,  \label{eq:dS}
\end{equation}

with $f(s,\epsilon )=f(z)$ the scaling function and $f^{\prime }(z)=\frac{df%
}{dz}$. If we evaluate equation (\ref{eq:dS}) at the critical point, i.e. 
$\epsilon =0$, we get:

\begin{equation}
\left(\frac{d S_1}{d \epsilon}\right)_{\epsilon=0} = \frac{\tau}{%
\zeta^2(\tau)} f^{\prime}(0) \sum_{s_1 s_2} s_1^{\sigma} s_1^{(-\tau)}
s_2^{(-\tau)} [log(s_1) -log(s_2)] \simeq -1.6 . \label{eq:d0}
\end{equation}

It is clear from this expression that the information entropy does not show
an extreme in the critical point for this family of distribution functions
and, therefore, $S_{1}$ is not a signal of critical behavior.

In Figure (2a) we also plot the $\chi^{2}$ coefficient for the cluster
distribution fitted with one parameter ($q_{0}=q_{0}(\tau)$ according to
equation (\ref{eq:q0})) and in figure (2b), the second moment of the distribution, 
$M_{2}$. It can be seen that these magnitudes signal the critical point very
accurately.

Finally, we apply these tools to molecular dynamics simulations of
fragmentation. These simulations have been described in previous works (see 
\cite{strador,balcrit,cherno,pre} for details). We follow the time evolution
of excited drops composed of 147 particles interacting via a Lennard-Jones
potential that undergo multifragmentation.

Using this model and taking the excitation energy as a control parameter we
perform the same analysis as in the case of the percolation model. The
corresponding results are displayed in figure (3). In this figure
we can see that $M_{2}$ and $NVM$ peak at $E=0.3\pm 0.2\epsilon $, meanwhile
the best fitted mass spectra according to equation (\ref{eq:q0}) (minimum $\chi^{2}$) 
is, approximately, at $E=0.15\pm 0.15\epsilon $. On the other hand the
information entropy monotonically increases in the range of energies
analyzed and therefore does not show a maximum when the system displays a
power law mass spectra, i.e. in the energies around $E=0.2\epsilon $.

In conclusion, we have analyzed different quantities proposed in the
literature as signatures of critical behavior: i) the information entropy 
$(S_{1})$, ii) the normalized variance of the size of the maximum fragment 
($NVM$) , iii) the second moment of the distribution ($M_{2}$), iv) the best
fitting power law and minimum $\tau $ . We found that $S_{1}$ is not a
reliable signature of a second order phase transition as was proposed in 
\cite{Ma}. For this purpose, we have analyzed systems that display critical
behavior: percolation on lattices of two sizes. For the larger one, where
the finite size effects can be neglected, it is clear that the signature
corresponding to information entropy is shifted with respect to the other
quantities analyzed. 
But when the size is much smaller and the finite size can not be neglected, 
the signatures analyzed in this work display peaks that do not coincide in a 
given value of the probability. Instead, they are smeared in a 
small range (see figures (1)). In this case, the broad peak of the entropy 
can turn out to be rather close to this range of values (even though the error is 
large) and then the information entropy can be confused with the signatures. 

This behavior is confirmed in the analysis of highly excited Lennard-Jones
drops that undergo multifragmentation.

Finally, according to these results, we would choose as signatures of
critical behavior the following signals: the best fit mass spectra with a
one-parameter power law (minimum $\chi ^{2}$), the normalized variance of
the size of the maximum fragment ($NVM$) or the second moment of the cluster
distribution ($M_{2}$) and we would discard the information entropy ($S_{1}$%
).

Acknowledgments : We acknowledge partial financial support from the
University of Buenos Aires via grant TW98 and Conicet via grant 4436/96.
C.O.D is member of the Carrera del Investigador (CONICET). P.B is fellow of
the CONICET .

\newpage

Figure Captions

Fig.1: The normalize variance of the size of the maximum fragment, $NVM$
(a), the second moment of the cluster distribution 
$M_2$ (b), the $\chi^2$ coefficient of a one parameter power law fitting 
(with $q_0=q_0(\tau)$) 
(c), the $\tau$ exponent from the fitting of two parameters power law (d)
and the information entropy ($S_1$) (e) against bond probability for
percolation lattices of sizes $30\times30\times30$ (full triangles) and 
$6\times6\times6$ (empty circles). In (f) we plot the cluster distribution 
for the critical probabilities at both sizes.

Fig.2: The $\chi^2$ coefficient (a) and $M_2$ (b) against probability for
the cluster distributions generated from equation (\ref{eq:fdm}) using the scaling
function $f(z)$ obtained in \cite{elliot_f}. In (c) we plot the information
entropy calculated by replacing the cluster distribution given by eq.(\ref
{eq:fdm}) in the definition of $S_1$ (equation (\ref{eq:S1})). In (d) we can see
both: the cluster distribution corresponding to the critical probability
(chosen at $p_i=0.25$, a pure power law) and the one that maximize the
information entropy ($p_i=0.22$). Here it is clear that the distribution that
maximize the entropy is not a power law.

Fig.3: The $\chi^2$ coefficient (a), the $M_2$ (b), the $NVM$ (c) and $S_1$
against the energy per particle for a molecular dynamics multifragmentation
process for 147 particles interacting via a Lennard-Jones potential.

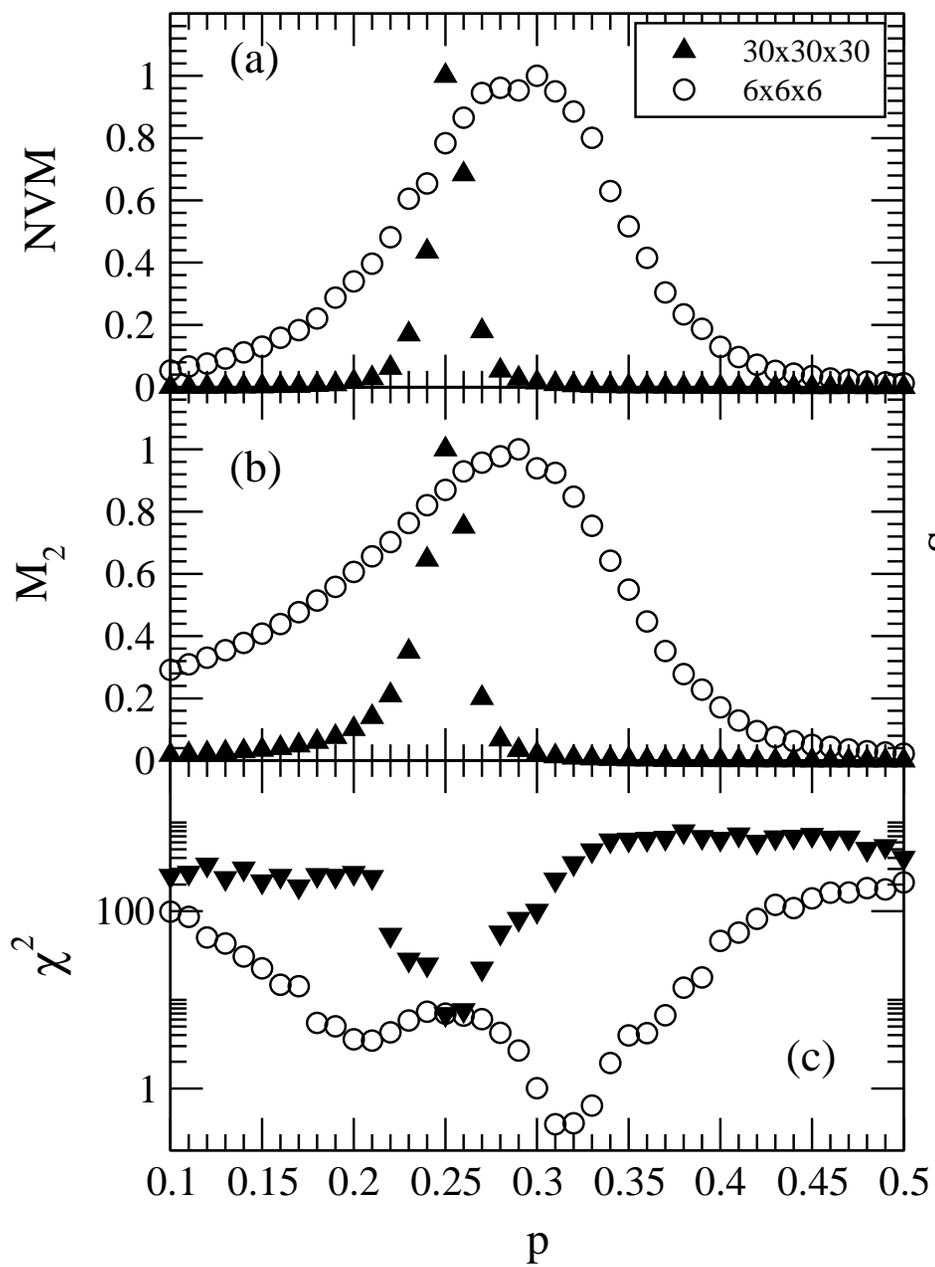
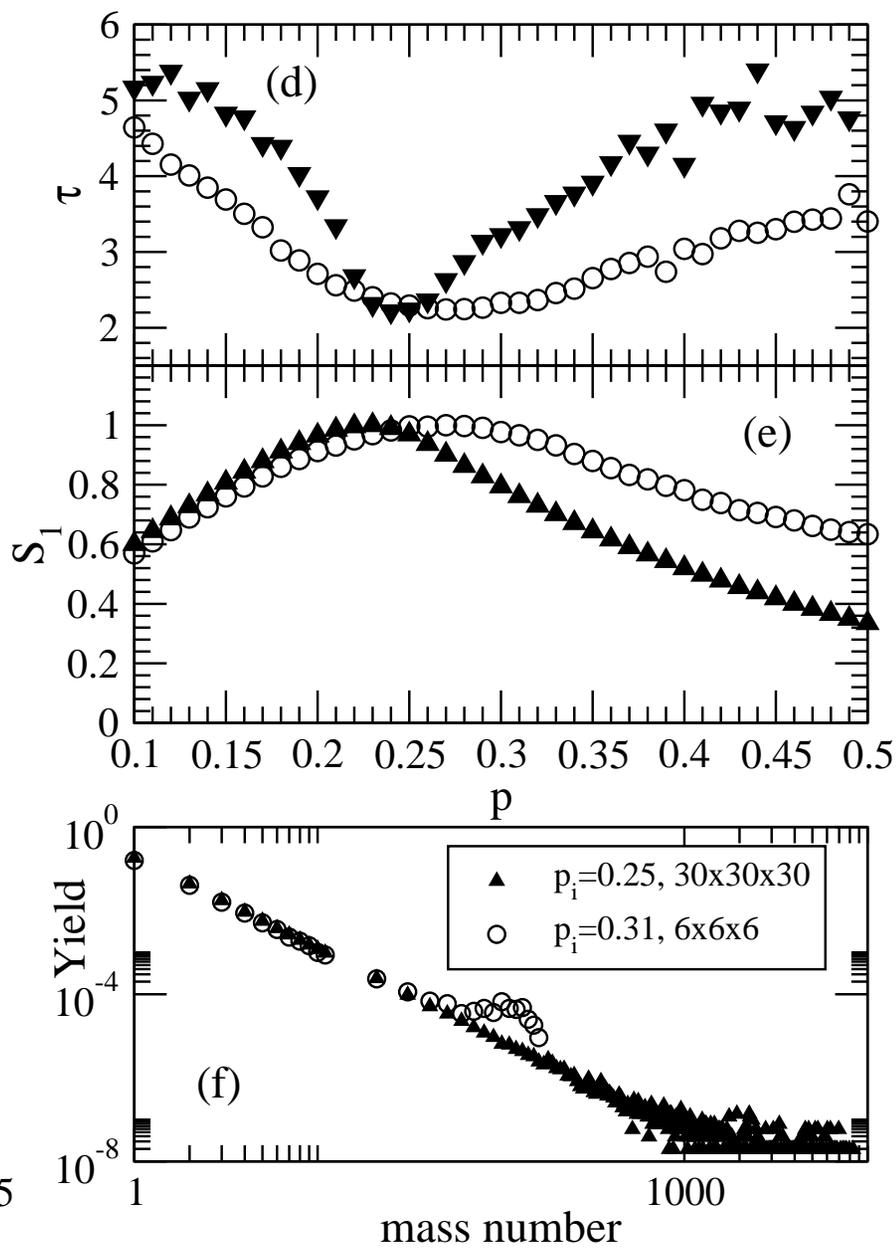
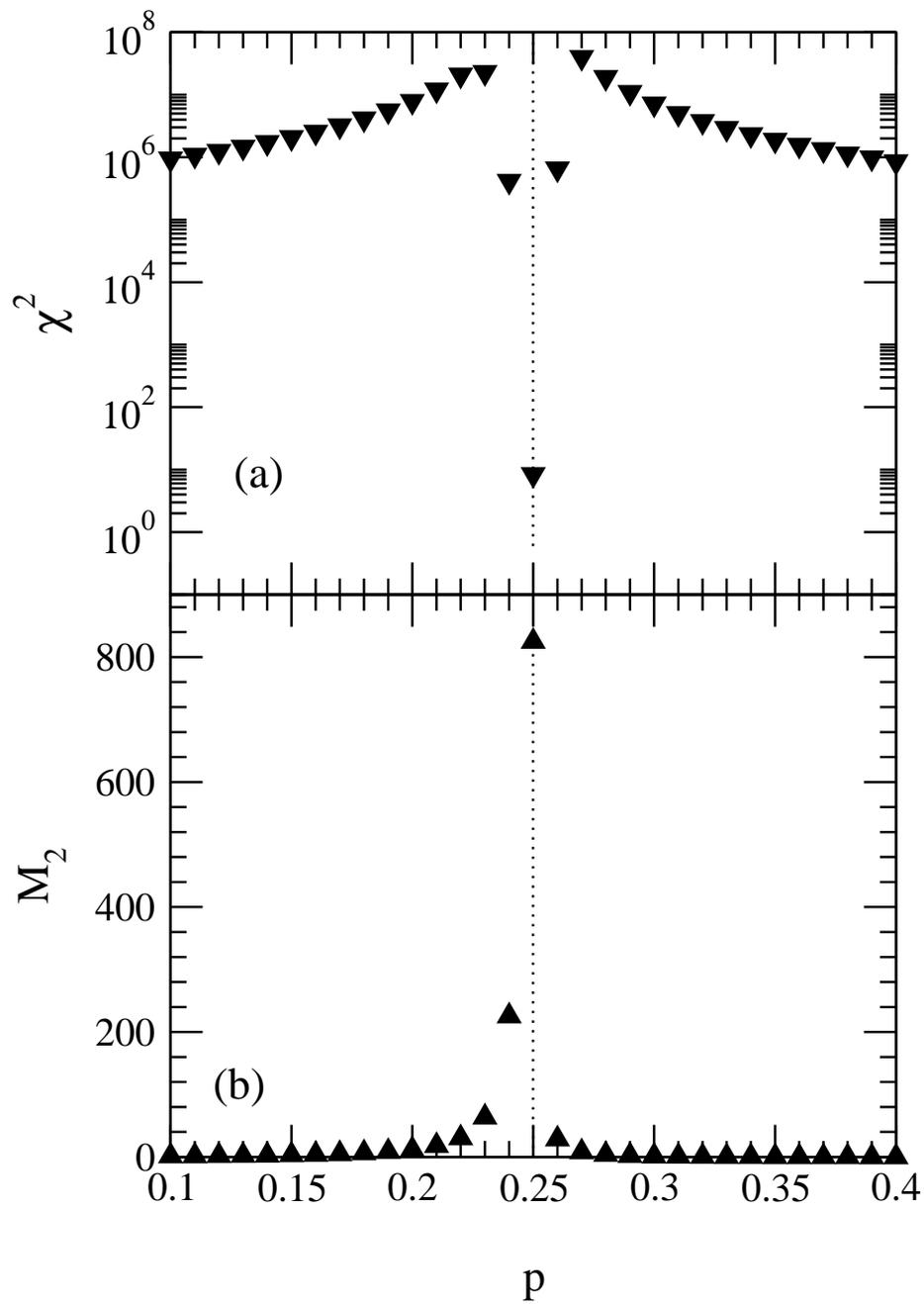
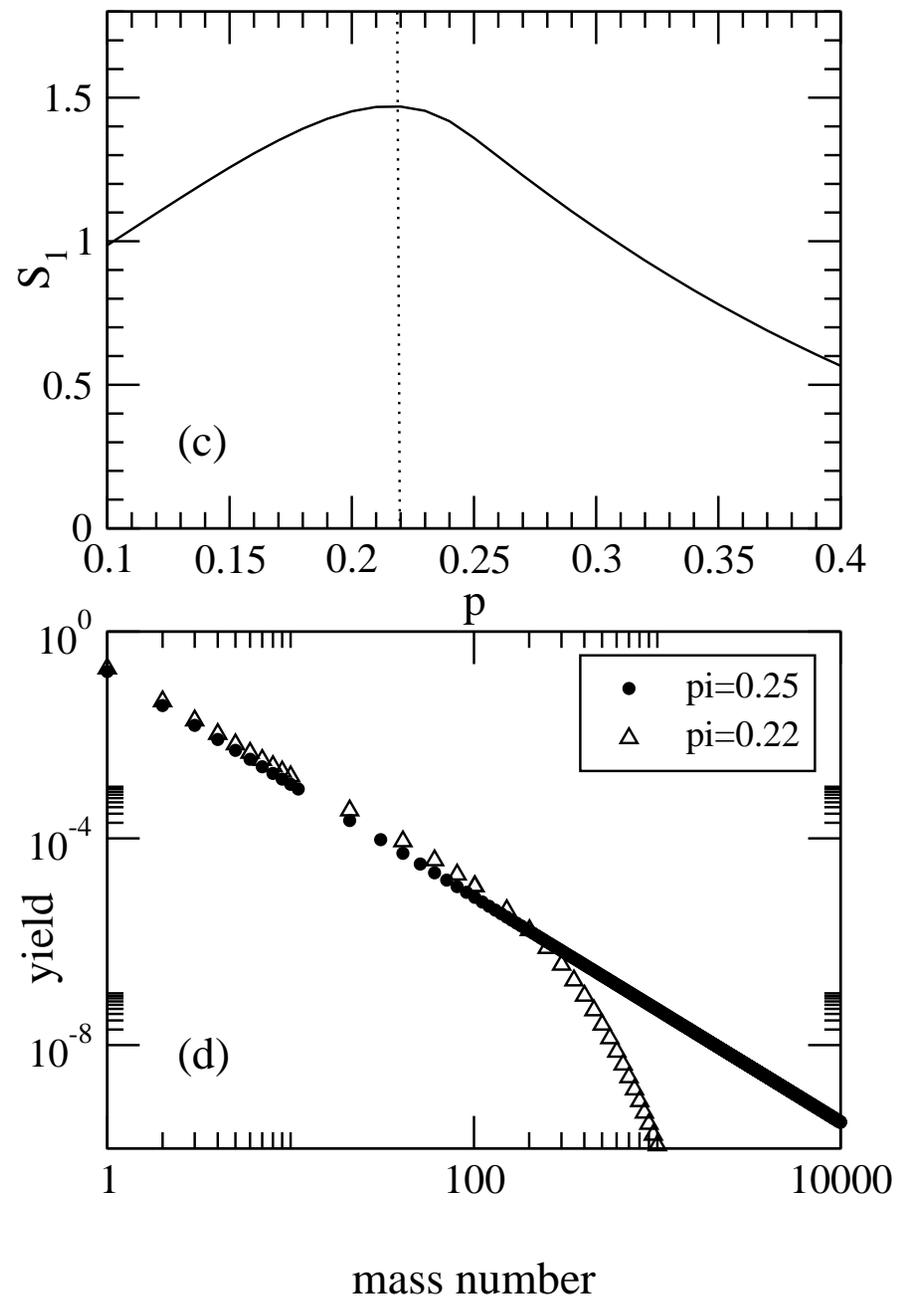
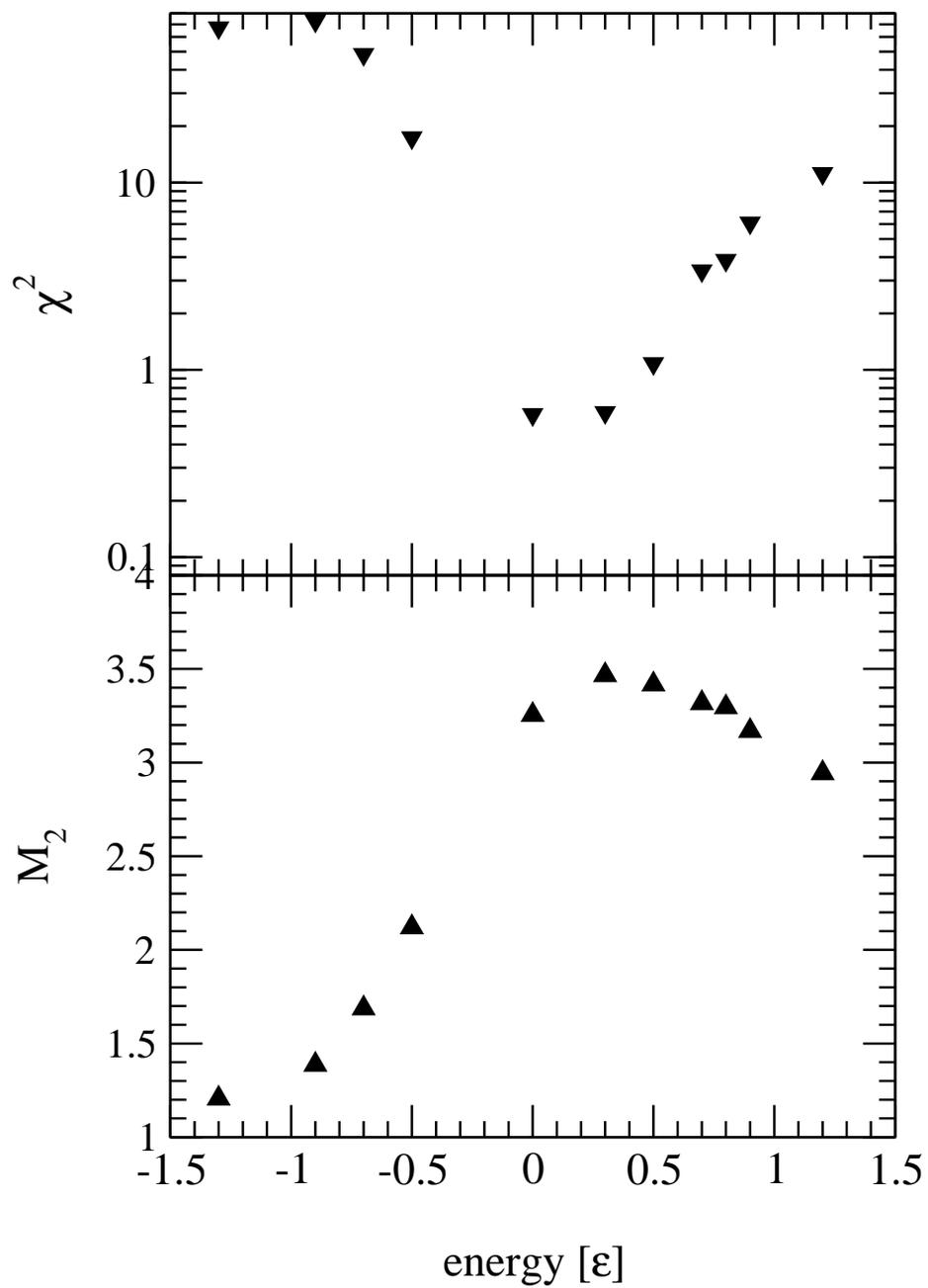
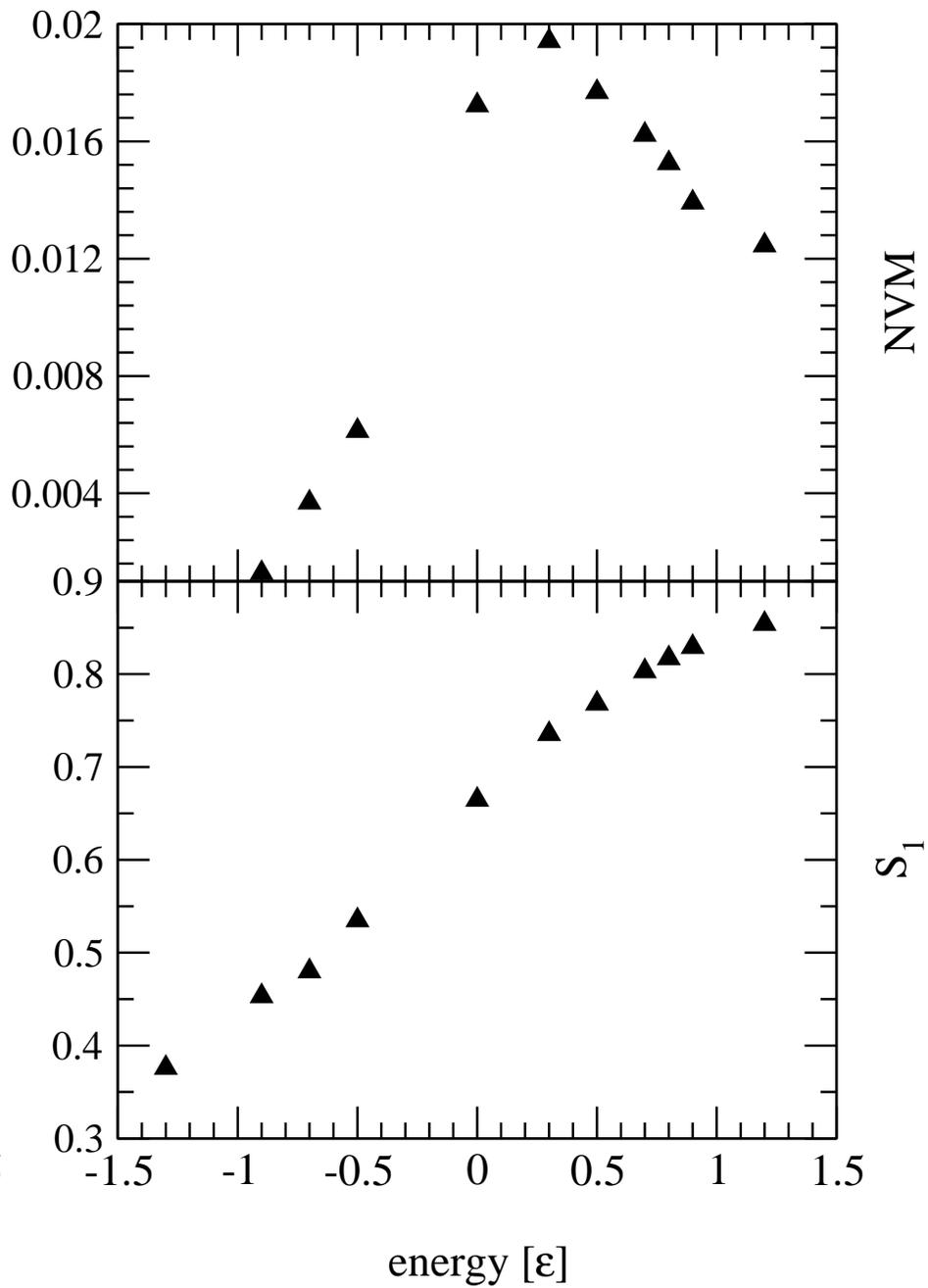
\end{document}